\documentclass[sigconf]{acmart}
\AtBeginDocument{%
  \providecommand\BibTeX{{%
    \normalfont B\kern-0.5em{\scshape i\kern-0.25em b}\kern-0.8em\TeX}}}

\copyrightyear{2023}
\acmYear{2023}
\setcopyright{rightsretained}
\acmConference[UbiComp/ISWC '23 Adjunct]{Adjunct Proceedings of the 2023 ACM International Joint Conference on Pervasive and Ubiquitous Computing & the 2023 ACM International Symposium on Wearable Computing}{October 8--12, 2023}{Cancun, Quintana Roo, Mexico}
\acmBooktitle{Adjunct Proceedings of the 2023 ACM International Joint Conference on Pervasive and Ubiquitous Computing \& the 2023 ACM International Symposium on Wearable Computing (UbiComp/ISWC '23 Adjunct ), October 8--12, 2023, Cancun, Quintana Roo, Mexico}\acmDOI{10.1145/3594739.3605107}
\acmISBN{979-8-4007-0200-6/23/10}

\usepackage{xcolor,color,soul,bm}
\usepackage{paralist}
\usepackage{hyperref}
\usepackage{multirow} 
\usepackage{tabularx}
\usepackage{balance}

\begin{document}

\title[FairComp '23]{FairComp: Workshop on Fairness and Robustness \\ in Machine Learning for Ubiquitous Computing}

\author{Sofia Yfantidou}
\email{syfantid@csd.auth.gr}
\orcid{0000-0002-5629-3493}
\affiliation{%
  \institution{Aristotle University of Thessaloniki}
  \city{Thessaloniki}
  \country{Greece}
}

\author{Dimitris Spathis}
\authornote{Also affiliated with the University of Cambridge, UK.}
\email{dimitrios.spathis}
\orcid{0000-0001-9761-951X}
\email{@nokia-bell-labs.com}
\affiliation{%
  \institution{Nokia Bell Labs}
  \city{Cambridge}
  \country{United Kingdom}
}

\author{Marios Constantinides}
\authornotemark[1]
\email{marios.constantinides}
\email{@nokia-bell-labs.com}
\orcid{0000-0003-1454-0641}
\affiliation{%
  \institution{Nokia Bell Labs}
  \city{Cambridge}
  \country{United Kingdom}
}
\orcid{}

\author{Tong Xia}
\email{tx229@cam.ac.uk}
\orcid{0000-0002-6994-6318}
\affiliation{%
  \institution{University of Cambridge}
  \city{Cambridge}
  \country{United Kingdom}
}
\author{Niels van Berkel}
\email{nielsvanberkel@cs.aau.dk}
\orcid{0000-0001-5106-7692}
\affiliation{%
  \institution{Aalborg University}
  \city{Aalborg}
  \country{Denmark}
}

\renewcommand{\shortauthors}{Yfantidou, et al.}

\begin{abstract}
How can we ensure that Ubiquitous Computing (UbiComp) research outcomes are both ethical and fair?
While fairness in machine learning (ML) has gained traction in recent years, fairness in UbiComp remains unexplored. This workshop aims to discuss fairness in UbiComp research and its social, technical, and legal implications. From a \emph{social perspective}, we will examine the relationship between fairness and UbiComp research and identify pathways to ensure that ubiquitous technologies do not cause harm or infringe on individual rights. From a \emph{technical perspective}, we will initiate a discussion on data practices to develop bias mitigation approaches tailored to UbiComp research. From a \emph{legal perspective}, we will examine how new policies shape our community's work and future research.
We aim to foster a vibrant community centered around the topic of responsible UbiComp, while also charting a clear path for future research endeavours in this field.

\end{abstract}




\maketitle

\vspace{-0.2cm}
\section{Background}
\label{sec:background}

Due to the integration of Machine Learning (ML) into Ubiquitous Computing (UbiComp), tasks that were once deemed impossible or reserved exclusively for humans, are now within technology's reach. Algorithms running on ubiquitous devices, such as smartphones and wearables, have been employed to recognize human activities~\cite{gu2021survey}, track sleep patterns~\cite{koskimaki2018we}, and detect breathing phases~\cite{10.1145/3369835}. Currently, we witness a surge in high-stakes applications such as diagnosing COVID-19 infections~\cite{10.1145/3394486.3412865}, detection of Atrial Fibrillation (AFib)~\cite{lubitz2022detection}, and enhancement of cognitive performance~\cite{costa2019boostmeup}. However, as with any technological advancement, ethical opportunities and risks come hand in hand, and, similarly to humans, ML algorithms are susceptible to biases.%

While fairness research has gained popularity in recent years \cite{10.1145/3457607,tahaei2023human}, with a dedicated conference and scientific community (FAccT), fairness in UbiComp remains wildly unexplored \cite{yfantidou2023review}. Yet, UbiComp applications are equally likely to suffer from biases. For example, health sensors such as oximeters consistently misclassify people of color~\cite{sjoding2020racial}, while client selection in federated learning incorporates biases against user profiles on wearables of inferior networking conditions~\cite{zhou2022you}.
Additionally, ubiquitous data and models have certain particularities, oftentimes not shared with the broader scholarly discourse on ML ethics. For example, they typically include small-scale, proof-of-concept datasets collected in the lab, making it difficult to extract population-level insights.
Also, they are mostly sequential/temporal in nature, with biases being harder to surface. In other words, while it is relatively straightforward to distinguish a person's skin tone from a picture, it is much harder to do so from sensor measurements; as UbiComp strives to blend technologies in the background, biases are blended too~\cite{constantinides2022good}.
Compounding the problem, the lack of geographical diversity, a dearth of longitudinal studies, and the under-reporting of essential study information create ``blind spots'' in the research on algorithmic fairness~\cite{van2023methodology}. 
Finally, given that such technologies are typically developed in Western countries, they tend to reflect the intuition, knowledge, and values of ``WEIRD'' (Western, Educated, Industrialized, Rich, and Democratic) cultures, potentially limiting their inclusivity and relevance to other contexts~\cite{septiandri2023weird,linxen2021weird,yfantidou2023review}.
However, with a conscious approach, it is possible to create models that are both robust and fair. As the field of ML continues to evolve, the UbiComp community needs to stay vigilant, ensuring that technological advancements are designed and deployed in a responsible and ethical manner.

To this end, we aim to spark a discussion about the ethical, social, technical, and legal issues relevant to fair and ethical UbiComp research. From a social perspective,  we look into how fairness research can be translated into this domain and identify pathways for ensuring that ubiquitous technologies do not cause any harm or infringe on any individual rights~\cite{constantinides2022good}. From a technical perspective, we intend to take a closer look at the community's data collection, processing, and modeling practices to ideate fairness enhancement and bias mitigation targeted at UbiComp work. From a regulatory perspective, we set out to understand how proposed policies, such as the European AI Act, will frame the work of the community and drive future research. This balance between performance and fairness is envisaged as a viable way forward---an ideal compromise.

These perspectives raise numerous challenges and questions that we seek to address in this workshop. 
How can we leverage existing fairness research and adapt it to the UbiComp domain? 
How can we define and quantify fairness in prevalent data (e.g., time-series) and model (e.g., regression, multi-class classification) modalities?
How can data and labels be acquired ethically?
How can we generate fair synthetic data or recruit representative real-world samples?
How do we incorporate fairness into our technology development processes and deployment monitoring by design?
Essentially, how do we better equip our community to deal with unfairness?

Besides these questions, which hold significance for the UbiComp community, several challenges extend to other disciplines, including Philosophy, Sociology, Law, Psychology, or any of the broad range of subjects contributing to this area. Which ethical challenges arise when technologies are interwoven into everyday life until they are indistinguishable from it? Which are the historical and systemic biases that frame the domain's research? How can ubiquitous applications be regulated without stifling innovation?

\section{FairComp Workshop}
\label{sec:workshop}

We aim FairComp to be an interdisciplinary forum beyond publications' solicitation that brings together academia and industry. Notably, we seek to bring together researchers and practitioners whose work lies within the ACM SIGCHI domains (e.g., UbiComp, HCI, CSCW), as well as FAccT, ML \& AI, Social sciences, Philosophy, Law, Psychology, and others. Workshop organizers are actively engaged in the aforementioned themes and will encourage their network of colleagues and students to participate in the workshop. In particular, the goal of this workshop is to collaboratively:
\begin{itemize}
    \item \textit{Assess} the evolving socio-technical themes and concerns in relation to fairness across ubiquitous technologies, ranging from health, behavioural, and emotion sensing to human-activity recognition, mobility, and navigation.
    \item \textit{Map} the space of ethical risks and possibilities regarding technological interventions (e.g., input modalities, learning paradigms, design choices).
    \item \textit{Envision} new sensing and data-acquisition paradigms to fairly and accurately gather ubiquitous physical, physiological, and experiential qualities.
    \item \textit{Explore} novel methods for bias mitigation and investigate their suitability for diverse ubiquitous case studies.
    \item More generally, \textit{start} a discourse around the future of ``ubiquitous fairness'' and co-create research agenda(s) for meaningfully addressing it. 
    \item Finally, \textit{consolidate} an international network of researchers to further develop these research agendas through funding proposals and through steering future funding instruments.
\end{itemize}

\noindent \textbf{Relevance and Impact to UbiComp.} With its strong community engaged in several themes (e.g., sensing, HCI, AI/ML), and its synergistic coalitions across varied domains (Sociology, Philosophy, Health Informatics, Law), UbiComp has a crucial role in paving the way for responsible, robust, and fair technological advancements. 
Coupled with the unique characteristics of ubiquitous technology, these advancements demand the development of distinct definitions, metrics, and methodologies to counteract the effects of bias.
This calls for the creation of a subcommunity focused on fairness issues in the domain's technology. By raising awareness and advocating decentralized work, we should encourage every member of the community to integrate fairness considerations in their research. Collaborating with other disciplines, this workshop aims to promote scientific exchange and jointly create a comprehensive and effective framework for ensuring fairness in UbiComp technology.

\vspace{3pt}
\noindent \textbf{Long-term Objectives.} This workshop will contribute to a deeper understanding of ethical challenges and opportunities surrounding the robust, and fair use of ubiquitous technology. 
Under such efforts, we plan to build an active and long-lasting community around the workshop's theme. Finally, we intend to use the workshop's momentum, as well as the developed research agendas and their collaborative follow-ups, to prepare a special issue of a journal (e.g., IEEE Pervasive) after the conclusion of the workshop. We plan to make an open call for this issue but will especially invite workshop participants to submit their work. Furthermore, we will consolidate our workshop's insights (including discussion) into an article.

\section{Workshop Structure}
\label{sec:workshop_structure}

\noindent\textbf{Workshop Topics.} The workshop aims to provide a platform for exchanging ideas that can shape the future of ubiquitous computing fairness and beyond and to rethink the role of UbiComp as an enabler of pervasive experiences free from biases.
The main topics of interest include, but are not limited to:

\begin{itemize}
    \item New definitions, metrics, and criteria of fairness and robustness, tailored for ubiquitous computing  %
    \item Indirect notions of fairness on devices (e.g., unfair resource allocation, energy, connectivity)
    \item New methods for bias identification and mitigation 
    \item Bias, discrimination, and measurement errors in data, labels, and under-represented input modalities
    \item New benchmark datasets for fairness and robustness evaluation (e.g., sensor data with protected attributes)
    \item Geographical equity across datasets and applications (e.g., WEIRD research, Global South)
    \item New user study methodologies beyond conventional protocols (e.g., Fairness-by-Design)
    \item Robustness (e.g., out-of-distribution generalization, uncertainty quantification) of ML models in high-stake and real-world applications
    \item Investigation of fairness trade-offs (e.g., fairness vs. accuracy, privacy, resource efficiency)
    \item Implications of regulatory frameworks for UbiComp
\end{itemize}

\noindent\textbf{Workshop Format.} We plan for an open, full-day workshop with 2 invited keynotes and 10 accepted papers that include completed and ongoing original empirical works, case studies, reviews, as well as position papers (subject to \# of submissions). All papers will be presented as talks, including Q\&A to allow researchers to engage in discussion with the workshop attendees. 
To further engage workshop participants, FairComp will include two interactive activities: 
a collaborative ideation session, where participants will be split into small groups to discuss the ethical, social, technical, and legal perspectives of UbiComp fairness under the guidance of invited experts and driven by questions provided by the organizers; and
an interactive panel on \emph{``Ethical \& Responsible UbiComp: A Case for Fairness and Robustness''} with keynote speakers and industry experts to further discuss reflections on their work with an open Q\&A session with the audience, moderated by one of the co-organizers. The workshop will take place on-site, with accommodation for exceptional cases' remote attendance via Zoom (\url{https://zoom.us/}). Additionally, Slack (\url{https://slack.com/}) will be used for facilitating social interaction. The entire workshop is estimated to last around 8 hours as illustrated in Table~\ref{tab:schedule}.

\begin{table}[t!]
\small
\vspace{-0.1pc}
\Description[Table showing proposed schedule]{Table showing the proposed schedule of the full-day virtual workshop.}
\caption{Proposed schedule for the FairComp workshop.}
\label{tab:schedule}
  
\begin{tabularx}{\columnwidth}{@{}llX@{}}
\toprule \\
    & \textbf{Time} & \textbf{Activity} \\
\midrule

\parbox[t]{3mm}{\multirow{7}{*}{\rotatebox[origin=c]{90}{\textbf{Session I}}}} 

& 09:00--09:15 & \textbf{Welcome:} Introduce organizers, participants, workshop objectives, and schedule. \\

& 09:15--10:15 & \textbf{Keynote \#1:} Presentation by an invited expert \newline (45-min talk followed by 15-min Q\&A). \\ 

& 10:15--10:30 & \textbf{Paper presentations \#1:} 2 paper presentations \newline (5-min talk followed by 2-min Q\&A). \\

& 10:30--10:45 & \textbf{Short Break}\\

& 10:45--12:00 & \textbf{Interactive Activity:} Collaborative ideation session in small groups about the ethical, social, technical, and legal perspectives of UbiComp fairness.  \\

\midrule

& 12:00--13:00 & Lunch Break \\

\midrule

\parbox[t]{5mm}{\multirow{10}{*}{\rotatebox[origin=c]{90}{\textbf{Session II}}}} 

& 13:00--14:00 &  \textbf{Keynote \#2:} Presentation by an invited expert \newline (45-min talk followed by 15-min Q\&A). \\ 

& 14:00--14:30 & \textbf{Paper presentations \#2:} 3 paper presentation \newline (5-min talk followed by 2-min Q\&A).   \\

& 14:30--14:45 & \textbf{Short Break}\\

& 14:45--15:30 & \textbf{Paper presentations \#3:} 5 paper presentation (5-min talk followed by 2-min Q\&A).   \\

& 15:30--15:45 & \textbf{Short Break}\\

& 15:45--16:30 & \textbf{Panel Discussion:} Keynote speakers and invited industry experts discuss the topic of ``Ethical \& Responsible UbiComp: a case for fairness and robustness''.  \\

& 16:30--17:00 & \textbf{Wrap Up:} Closing remarks and best paper award.\\

     \bottomrule
  \end{tabularx}
\end{table}

Session I will begin with the first keynote. We contacted Prof. Flora Salim (University of New South Wales, Australia), who kindly agreed to share lessons about fairness and robustness on UbiComp. The first half will continue with paper presentations. It will conclude with an interactive activity aimed at sparking a discussion around UbiComp fairness perspectives among the participants while building future collaborations. 
Session II will start with the second keynote. Prof. Ricardo Baeza-Yates (Institute for Experiential AI of Northeastern University, USA) agreed to talk about computational fairness and human-centric computing. The remaining paper presentations will follow, along with a panel discussion (Dr Akhil Mathur from Meta AI agreed to be the industry expert panellist) on the topic to consolidate ideas into an executable research agenda.
The workshop will wrap up with the best paper award.

\vspace{3pt}
\noindent\textbf{Estimated Number of Participants.}
The growing interest in the workshop's theme is demonstrated by the proliferation of similar workshops at other prestigious conferences, including FairUMAP at UMAP, Trustworthy and Socially Responsible ML (TSRML) at NeurIPS, and Trustworthy ML in Healthcare at ICLR. Yet, none of these workshops focused on the particularities of ubiquitous technologies, nor did they target the UbiComp community. 
We consider 30-40 attendees an appropriate size for our workshop, allowing us to shape a comprehensive future research agenda, build collaborations, and consolidate an active community around the workshop theme. 

\vspace{3pt}
\noindent\textbf{Paper Selection and Publication.}
Submissions will be reviewed by at least two or three reviewers (including organizers and external reviewers) and will be published by ACM.
Our acceptance criteria will be a mixture of relevance, novelty, provocativeness, and research quality. Given the timely theme of this workshop, we are confident of attracting a large number of paper submissions which will enable us to organise a high-quality workshop.

\vspace{3pt}
\noindent\textbf{Important Dates.}

\begin{compactitem}
\item Call for Papers (CfP): 18 April 2023
\item Paper submission: 23 June 2023
\item Notification to authors: 8 July 2023
\item Camera-ready deadline: 31 July 2023
\item Workshop Day: 9 October 2023 
\end{compactitem}

\vspace{3pt}
\noindent\textbf{Pre-workshop Activities.}
We will disseminate the call for papers (CfP) through diverse channels, including mailing lists, our social and professional networks, local ACM chapters, the dedicated workshop website, and our respective institutional communication channels.
The website (\url{https://faircomp-workshop.github.io/2023/}) will be a key platform to disseminate information, including the CfP, crucial deadlines, profiles of the co-organizers, Technical Program Committee (TPC), workshop schedule, and activities. Moreover, the website will also serve as an archive of the workshop outcomes, containing the workshop's summary and other outputs.

Upon acceptance, we will reach out to experts from academia and industry to compose a TPC to review and select author contributions and facilitate preparing a diverse and thematically-rich program. In constituting the TPC, we will aim to find a balance among the themes relevant to the workshop. We will invite submissions of different kinds, ranging from technical papers, work-in-progress, and reviews to position papers, provocations, and case studies. The submissions will be 4--6 pages long, including references, and organizers will provide a template on the website.

\vspace{3pt}
\noindent\textbf{Post-workshop Activities.}
During and following the workshop, the results and outcomes will be blogged on the workshop website and disseminated in ACM Interactions. Drawing on the workshop submissions and interactive activities, we will propose a journal special issue (e.g., IEEE Pervasive) and encourage participants to collaborate on submissions around the developed research agendas. 

\vspace{3pt}
\noindent\textbf{Diversity Statement and Accessibility.}
Our workshop is committed to creating a welcoming and inclusive environment for all attendees, regardless of race, gender, sexual orientation, religion, or ability. This belief will be implemented by ensuring the diverse selection of organizers, TPC members, and speakers, disseminating the CfP in mailing lists targetting under-represented communities in computing, promoting inclusive language, and disseminating the conference's code of conduct. 
We aim to make our workshop inclusive to diverse participants, including access to materials. We plan on ensuring accessibility throughout the workshop. Prior to submission, the authors will be asked to adhere to UbiComp's Accessible Submission Guidelines.\footnote{\url{https://www.ubicomp.org/ubicomp-iswc-2023/accessibility/accessibility-guidelines/}} 
Finally, in the weeks leading up to the workshop, we will conduct a survey with attendees to identify the accessibility needs for in-person participation to accommodate during the workshop in collaboration with the JEDI Chairs.

\section{Organizers}

\label{sec:organizers}

\noindent\textbf{Sofia Yfantidou} is an early-stage researcher at the Aristotle University of Thessaloniki, and a Marie Skłodowska-Curie fellow at the Innovative Training Network ``Real-time Analytics for the Internet of Sports''. She works at the intersection of UbiComp and ML fairness.  Her current research focuses on defining, quantifying, and mitigating biases in data and ML models for health and well-being. She is a Heidelberg Laureate Forum alumna and a Grace Hopper scholar.
\texttt{\textbf{Website:}} 
\url{https://www.linkedin.com/in/sofiayfantidou/} 
\smallskip

\noindent\textbf{Dimitris Spathis} is a research scientist at Nokia Bell Labs, Cambridge (UK) and a visiting researcher at the University of Cambridge. His work enables AI to make the most out of real-world multimodal and sequential data through label-efficient and robust ML. He previously worked at Microsoft Research, Telefonica Research, Ocado, and Qustodio. His experience as an organizer of scientific meetings includes WellComp at Ubicomp '22, CHIL '23, Federated Sensing tutorial at MobiCom '21, and ML4H co-located with NeurIPS '22.
\texttt{\textbf{Website:}} 
\url{https://www.cl.cam.ac.uk/~ds806/} 
\smallskip

\noindent\textbf{Marios Constantinides} is a Senior Research Scientist at Nokia Bell Labs, Cambridge (UK) and a visiting researcher at the University of Cambridge. He works in the areas of human-computer interaction,  UbiComp, and responsible AI. His current research focuses on building AI-based technologies that augment people's interactions and communication, with a particular focus on the workplace. He has been a member of the organizing committee of the SensiBlend workshop at UbiComp '21, and co-organized two Special Interests Groups (SIG) at CHI '23 on future of work and responsible AI.
\texttt{\textbf{Website:}} 
\url{https://comarios.com/} 
\smallskip

\noindent\textbf{Tong Xia} is a third-year PhD candidate 
at the University of Cambridge. Her research interests lie in data mining, ML, and UbiComp for public health and human well-being. Particularly, she is keen to develop data-efficient, high-performance, uncertainty-aware, and privacy-preserving mobile health systems. 
She previously worked at Tecent, and she has been a committee member of the UK-Tsinghua association.  She also served as the Posters\&Demos session chair in UbiComp '22. 
\texttt{\textbf{Website:}} 
\url{https://xtxiatong.github.io}
\smallskip

\noindent\textbf{Niels van Berkel} is an Associate Professor at Aalborg University. His work focuses on the design and evaluation of intelligent computing systems, particularly in real-world contexts, publishing in HCI, Social Computing, and Ubiquitous Computing. He has previously served as organiser of workshops at UbiComp (UbiTtention'20, Mobile Human Contributions '18, Sensors \& Behaviour '18) and CHI (2VT '21, Emergent Interaction '21), and served on the editorial board for IJHCS (2019--present) and ACM TiiS Special Issue on Human-Centered Explainable AI.
\texttt{\textbf{Website:}} 
\url{https://www.nielsvanberkel.com/}

\balance

\bibliographystyle{ACM-Reference-Format}
\bibliography{main}

\end{document}